\documentclass[twocolumn]{aastex63}

\usepackage[caption=false]{subfig}
\usepackage{fontawesome}
\usepackage{CJK}

\usepackage{graphicx}
\begin{document}
\begin{CJK*}{UTF8}{gbsn}
\title{Searching For Transiting Planets Around Halo Stars. II. Constraining the Occurrence Rate of Hot Jupiters}

\email{boley.62@osu.edu}

\author[0000-0001-8153-639X]{Kiersten M. Boley}

\author[0000-0002-4361-8885]{Ji Wang (王吉)}
\affiliation{Department of Astronomy, The Ohio State University, Columbus, OH 43210, USA}

\author[0000-0002-7550-7151]{Joel C. Zinn}
\altaffiliation{NSF Astronomy and Astrophysics Postdoctoral Fellow}
\affiliation{Department of Astronomy, The Ohio State University, Columbus, OH 43210, USA}
\affiliation{Department of Astrophysics, American Museum of Natural History, Central Park West at 79th Street, NY 10024, USA}

\author[0000-0001-6588-9574]{Karen A.\ Collins}
\affiliation{Center for Astrophysics \textbar \ Harvard \& Smithsonian, 60 Garden Street, Cambridge, MA 02138, USA}

\author[0000-0003-2781-3207]{Kevin I.\ Collins}
\affiliation{George Mason University, 4400 University Drive, Fairfax, VA, 22030 USA}

\author[0000-0002-4503-9705]{Tianjun Gan}
\affiliation{Department of Astronomy and Tsinghua Centre for Astrophysics, Tsinghua University, Beijing 100084, China}

\author[0000-0002-9110-6163]{Ting~S.~Li}
\altaffiliation{NHFP Einstein Fellow}
\affiliation{Observatories of the Carnegie Institution for Science, 813 Santa Barbara St., Pasadena, CA 91101, USA}
\affiliation{Department of Astrophysical Sciences, Princeton University, Princeton, NJ 08544, USA}





\begin{abstract}

Jovian planet formation has been shown to be strongly correlated with host star metallicity, which is thought to be a proxy for disk solids. Observationally, previous works have indicated that jovian planets preferentially form around stars with solar and super solar metallicities. Given these findings, it is challenging to form planets within metal-poor environments, particularly for hot Jupiters that are thought to form via metallicity-dependent core accretion. Although previous studies have conducted planet searches for hot Jupiters around metal-poor stars, they have been limited due to small sample sizes, which are a result of a lack of high-quality data making hot Jupiter occurrence within the metal-poor regime difficult to constrain until now. We use a large sample of halo stars observed by TESS to constrain the upper limit of hot Jupiter occurrence within the metal-poor regime (-2.0 $\leq$ [Fe/H] $\leq$ -0.6). Placing the most stringent upper limit on hot Jupiter occurrence, we find the mean 1-$\sigma$ upper limit to be 0.18 $\%$ for radii 0.8 -2 R$_{\rm{Jupiter}}$ and periods $0.5- 10$ days. This result is consistent with previous predictions indicating that there exists a certain metallicity below which no planets can form.


\end{abstract}
\section{Introduction} 

The evolution of the Milky Way and planet formation are connected \citep{Forgan2017}. Metallicity links these seemingly unrelated topics by playing an important role in the formation and evolution of stars and therefore affects the formation of giant planets \citep{Johnson2009,Choi2009}. Host star metallicity is thought to reflect the metallicity of the protoplanetary disk from which planets form \citep{Wyatt2007,Haworth2016}. Given that the first generation of stars was metal-poor, the first planets must also be lacking in metals. By probing planet occurrence in the metal-poor regime, we can gain insight into the first generation of giant planet formation and determine the metallicity at which no planet can form.

A metal-poor environment poses many challenges for planet formation. At low metallicities, the protoplanetary disk lifetime is significantly shorter which decreases the likelihood of planet formation \citep{Kornet2005, Yasui2010, Ercolano2010}. The vast majority of the gas within the disk is hydrogen and helium, and heavy elements compose only a fraction  $\sim 10^{-5}$ of the total mass in the metal-poor regime \citep{Johnson2012}. A shortened disk lifetime can be detrimental to planet formation, especially at short orbital distances where planet formation is thought to be dominated by core accretion \citep{Miller2011,Bailey2018}. Although we know that planet formation is increasingly improbable with decreasing metallicity, the metallicity for which no planet can form is still uncertain.

Since the discovery of four Jovian planets orbiting metal-rich stars in 1997 by \cite{gon1997}, the correlation between planet occurrence and host star metallicity has been 
a point of focus within the exoplanet community. Following this discovery, various studies also noted that Jovian planets preferentially orbit metal-rich stars \citep{Santos2004, Fischer2005, mortier13}. The occurrence of Jovian planets has been shown to increase exponentially with the increase of metallicity for solar and super-solar metallicities \citep{Udry2007,Johnson2010,mortier13}; however, the hot Jupiter (HJ) occurrence trend within the metal-poor regime is still uncertain. Using radial velocities, \cite{sozzetti2009} and \cite{mortier2012} probed this regime to determine Jovian planet occurrence and provide an upper limit for HJ occurrence in the metal-poor regime. Since the study conducted by \cite{mortier2012}, there has been little investigation to further constrain HJ planet occurrence at low metallicities.

With the advent of Transiting Exoplanet Survey Satellite~\citep[TESS, ][]{TESS}, it is the opportune time to revisit HJ occurrence in the metal-poor regime as TESS provides a crucial advantage compared to Kepler to complete this work. The major advantage of
TESS is its larger field of view, which allows for 30-minute cadence Full Frame Images (FFI). These FFIs provide observations of a wide range of stellar types. TESS also covers $\sim 85\%$ of the sky, which is a significant improvement to the $\sim 0.25\%$ monitored by Kepler. The capabilities of TESS have enabled a more robust determination of the HJ occurrence within the metal-poor regime which was not possible until now. 

In this paper, we focus on further constraining HJ planet occurrence in the metal-poor regime (-2.0 $\leq$ [Fe/H] $\leq$ -0.6) using a population of halo stars, as they provide important insights into planet-formation processes in the first generation of stars within the Milky Way. Our approach to deriving HJ occurrence rates is largely inspired by the method first outlined by \cite{Dressing2015}. We perform a full transit search of our halo star sample, using both the Box Least Squares algorithm \citep{Kovacs2002} and Transit Least Squares algorithm \citep{tls2009} to search the light curves. Using an injection-recovery scheme to calculate our detection efficiency, we use our results to constrain the occurrence rate of HJs orbiting halo stars.

The outline of our paper is as follows. Section \ref{s:stellar} provides a description of our stellar sample. In Section \ref{s:acquire}, we discuss the method from which we acquire TESS data. We explain our planet detection pipeline in Section \ref{s:detection}.  Section \ref{s:injection} contains a description of our planet injection pipeline. In Section \ref{s:completeness}, we assess the completeness of our pipeline. Section \ref{s:occur} is dedicated to a statistical analysis of the data and estimation of the planet occurrence rate. We compare our results to previous studies in Section \ref{s:comparison} and their implications for the frequency of HJs in the metal-poor regime before concluding in Section \ref{s:summary and conclusions}.

\begin{table}[t]
\label{tab:stellar}
\caption{Stellar Sample Parameters}
\centering
\begin{tabular}{cccc}
\hline
Parameter & Range & Median & Units\\
\hline
\hline
Mass & 0.34-1.29 & 1.02&$M_\odot$\\
Radius & 0.35- 1.17 &0.71&$R_\odot$\\
log(g) & 4.37-4.94&4.73& cgs\\
TESS Magnitude & 12.36-15.59 &14.25& mag\\
Effective Temperature & 3780-6440 &5714&K\\
\hline
\end{tabular}
\tablecomments{We use the 5$\%$ to 95$\%$ quantiles for each parameter.}
\end{table}

\section{Stellar Sample} \label{s:stellar}
\begin{figure}
\begin{center}
\includegraphics[width=\linewidth]{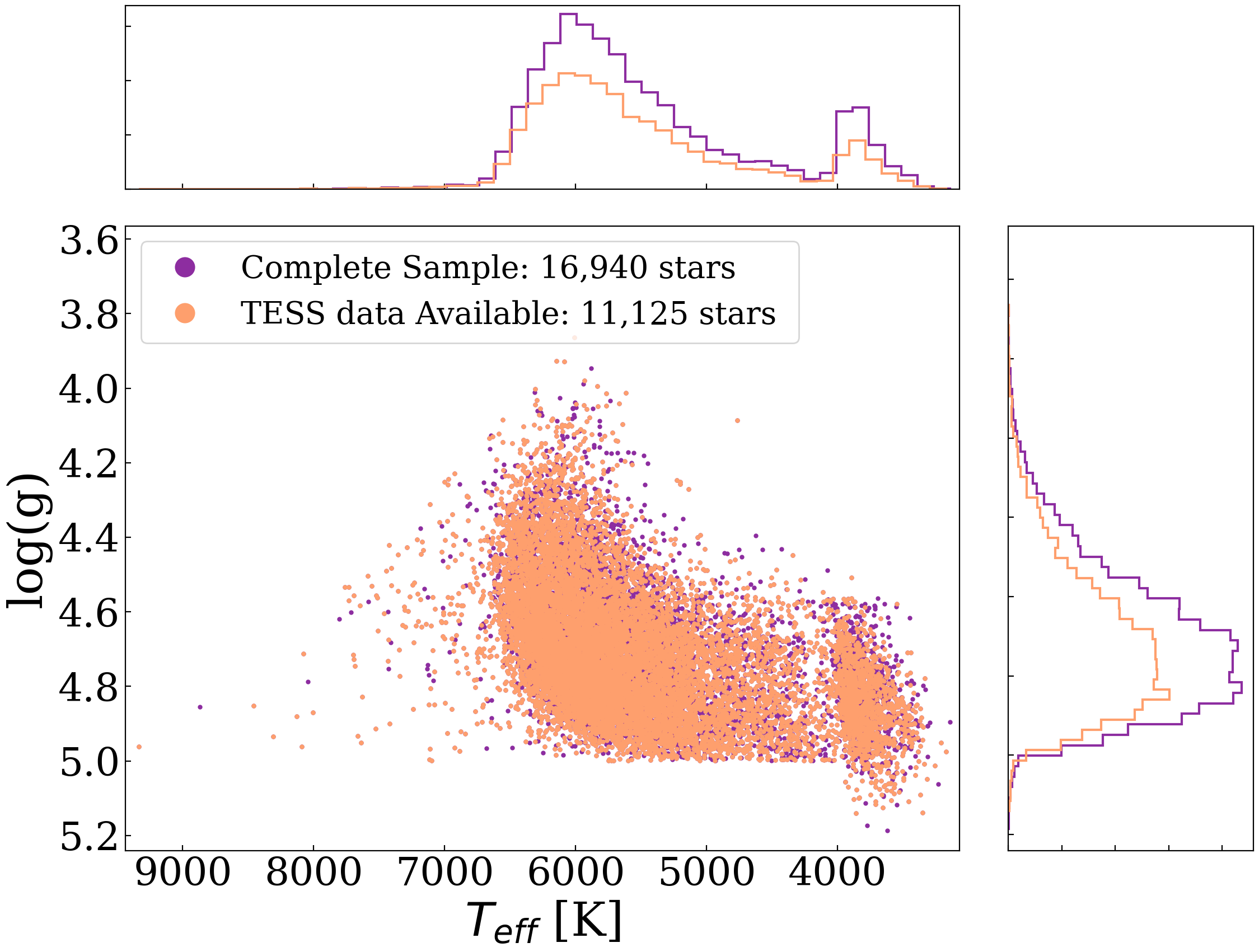}
\caption{Surface gravity and stellar effective temperatures of the stars obtained via the TESS Input Catalog~\citep{Statsun2018} in
our final sample (cyan)  compared to the complete sample (purple)
initially selected based on kinematic information. The bump at high surface gravity and low effective temperature most likely results from these stars being a cool dwarfs. \label{fig:stellardist}}
\end{center}
\end{figure}

We selected $\sim$16,940 halo stars, which are typically sub-dwarfs, to search for HJs using TESS data. Details on the sample selection and validation can be found in a companion paper (Kolecki et al. 2021); however, we briefly describe the sample selection here. The sample is selected using the transverse kinematic information, which was constructed following \citet{Koppelman2018}. Using radial velocities and proper motions from Gaia DR2 \citep{Gaia2018}, we select stars within 1 kpc and with 2D velocities greater than 210 km/s with respect to the local standard of rest. The sample was chosen using heuristic tangential motion cuts based on simulations of the Milky Way from Galaxia~\citep{Sharma2011}, which is outlined in Kolecki et al. 2021. By conducting a population study of $\sim$120 stars in our sample with APOGEE spectra (Kolecki et al. 2021), we confirmed that 70$\%$ of the selected targets are genuine halo stars. Table \ref{tab:stellar} list the ranges of various parameters of the sample, which were obtained via the TESS Input Catalog \citep{Stassun2018}. An H-R diagram of the sample is also provided in Fig. \ref{fig:stellardist}. The discontinuity in the distribution is due to the in-homogeneous treatment to low-mass stars and other stars in the TESS Input Catalog~\citep[see Appendix E.1.1 in][]{Stassun2018}. 

From the initial sample, we obtained 30-min-cadence data for $\sim 65\%$ (11,125 out of 16,940) from TESS through sector 23 \citep{TESS} Given that only a small fraction ($\sim 1\%$) of the sample had TESS 2-min cadence data, we chose to take advantage of a majority of the sample using the 30-min-cadence data. These light curves were obtained from TESS Full Frame Images (FFI) with \texttt{Eleanor}~\citep{2019PASP..131i4502F}. The FFI light curves are then used for planet detection and injection-recovery tests. 

\section{Acquiring TESS data} \label{s:acquire}
\texttt{Eleanor}~\citep{2019PASP..131i4502F}, software that was designed for downloading and analyzing TESS FFI data, is employed to acquire the 30-min-cadence TESS FFIs for our sample. Using the TESS magnitude of each target, we determine the optimal aperture size, which is then input into Eleanor. We download the four different data files: “PCA”, “Correlated”, “Raw”, and “PSF”. 
From there, we determine the noise level of each light curve. We normalize the flux data by dividing the light curve by the median flux. The noise of each file is determined by calculating the RMS noise of the light curve. The data file with the least amount of noise is saved.

\section{Planet Detection Pipeline and Results} \label{s:detection}

To determine the occurrence rates and  completeness of our search, we must search within our sample for planet candidates.  We developed an automated pipeline that encompasses all the data processing to include detrending, noise removal, and  transit search. Given that our target stars are relatively faint, we must minimize the noise where possible. From acquiring the FFI data, we choose the light curve with the least about of noise. We also apply the most robust detrending method available within the detrending program used. 

We begin by detrending the light curves. The data is initially detrended using \texttt{Wotan} ~\citep{wotan2019} with the Tukey's bi-weight method, which is indicated to be the most robust detrending method within their work \citep{wotan2019}. Since the TESS data has sharp peaks at the beginning and ends of an observation which we consider to be “data gaps”, we perform sigma-clipping to remove any systematics that \texttt{Wotan} was unable to remove. We determine the outliers at the edges of the light curves and at the edges of the data gaps. We only consider 5$\%$ of the data points on the edges when conducting the sigma clipping as they most negatively impact the transit search algorithms. We then calculate the median of the complete light curve and any points that are over 3$\sigma$ within the range considered to be the edges of the data are removed. From here our pipeline splits into two separate transit search algorithms: Transit Least Squares~\citep[TLS, ][]{tls2009}) and  Box Least Squares algorithm~\citep[BLS, ][]{Kovacs2002}). 

TLS, an open source transit search python package, is noted to be more efficient at conducting transit searches for small planets \citep{tls2009}. To expand upon their work, we chose to test whether it also performs better than BLS for giant planets around halo stars. To test how well TLS performs, we used the default transit template. For each star, we also input the limb darkening coefficients, mass, and radius to optimize for our sample.

For our BLS algorithm, we used the reference implementation of BLS provided by \texttt{astropy 4.0.1} \citep{Astropy2013}. With this implementation the false alarm probability (FAP) and signal to noise ratio (SNR) are not provided; however, they are required within the vetting process. In Section \ref{s:bls}, we expand upon how we calculate the FAP, and SNR given that these values are not calculated by the BLS program that we employed.

\subsection{BLS} \label{s:bls}
Given our choice to use two transit search algorithms, we make every effort to conduct a fair comparison. Following \cite{tls2009}, we empirically determined the log-likelihood for an FAP value of 0.01$\%$. Similar to TLS, we generated 10,000 synthetic light curves with only white noise with a 1-hour Combined Differential Photometric Precision of 290 ppm. We choose this noise level to replicate the typical noise level for TESS for stars with a TESS magnitude of 12 \citep{TESS}, which is the mean TESS magnitude for our sample. It is important to note that TESS light curves have correlated noise resulting from instrumental trends; however, we use white noise to ensure an equal comparison between TLS and BLS. Once the white noise only light curves have been evaluated by BLS, we impose an FAP value of 0.01 $\%$ determining the corresponding log-likelihood threshold for BLS.

Within BLS, the signal-to-noise ratio (SNR) is not automatically determined; therefore, we calculate a SNR for each fit as
\begin{equation}
 SNR= \frac{d}{\sigma}\sqrt{n_{transit}}
 \end{equation}
where \textit{d} is the difference between the median of the data points in the transit and the flux level outside the transit, $n_{transit}$ is the number of points within
transit, and $\sigma$ is the the standard deviation of the detrended light curve.

\subsection{Edi-vetter Unplugged}

Within our planet detection pipeline, we employ \texttt{Edi-vetter Unplugged} \citep{edi} to supplement our vetting tests. \texttt{Edi-vetter Unplugged} is an extension of Edi-vetter, which was designed to run in sequence with TLS to automatically vet planet candidates. Within this vetting algorithm, multiple tests are performed to remove false positives, which are briefly described as follows. 

\begin{itemize}
  \item Flux contamination test: determines whether a significant amount of flux from a nearby star may have blended with the pixels of the target.
  \item Outlier test: tests aimed to identify outliers that may align causing a false positive.
  \item Individual transit test: calculates the Bayesian
Information Criterion for an single transit to ensure the expected transit signal strength is large enough to provide a meaningful fit.
\item Even/odd mismatch test: identifies if the transit
primary and secondary eclipse are folded on top of each
other.
\item Uniqueness test: determines cases where the phase folded
light curve appears to produce several transit-like dips.
\item Secondary eclipse test: tests the secondary
eclipse model depth to determine if it is greater than 10 $\%$ of the depth of the initial transit and therefore statistically significant. 
\item Phase coverage test: when removing bad quality data from light curves, the data removed can line up in phase space causing a false positive. This test determines whether the transit contains sufficient data
to detect a statistically significant transit event.
\item Transit duration limit test: a check to determine
whether the detected transit duration is too long for the
detected transit period.
\item False positive test: after all false positive tests are complete, this flag indicates whether the planet is determined to be a false positive. If any of the previous test are determined to be a false positive, this flag will be triggered. 
\end{itemize}

\subsection{Vetting } \label{s:vetting}

Using our transit detection pipeline, we identified 2128 transit events with SNR greater than 5$\sigma$. Some of those signals might have been systematics or astrophysical false positives instead of transiting planet candidates. Therefore, we conducted a series of cuts to select the events consistent with transiting planets, which is described below.

After the light curves are searched using both TLS and BLS, any potential detections we require a FAP value $\leq$ 0.0001. The remaining candidates were then sent through \texttt{Edi-vetter Unplugged} \citep{edi} to vet the light curves. If the candidates did not pass all 8 tests within Edi-vetter Unplugged, they were removed. On the remaining candidates, we conducted visual examinations. We checked for the appearance of secondary eclipses and considered the depth of the transit relative to other potential features. We also checked for outliers that were not detected by Edi-vetter. From the visual examinations, 2 planet candidates remained. The rejected signals were determined to be either eclipsing binaries or systematics.

\subsection{Follow-up observations \& Results}

The two candidates that survived our vetting process and continued on to follow-up observations are TIC 229802010 and TIC 188593930. For TIC 229802010, we took one high-resolution spectrum using the MIKE spectrograph on the Magellan telescope~\citep{Bernstein03} on UT Jul 27, 2019 (PI: Ting Li). This spectrum had an exposure time of 1500 seconds resulting in an SNR of 10 per pixel; therefore, providing a sufficient SNR to cross-correlate a solar-type synthetic spectrum with the observed spectrum. The cross-correlation function clearly showed double peaks that were separated by $\sim$50 km/s, which is indicative of a grazing eclipsing binary (EB) system. 

Within the TESS follow-up observation collaboration, we took two spectra with CHIRON~\citep{Tokovinin2013}, resulting in two RV points at a phase of approximately 0.45 and 0.82 for TIC 188593930 (PI: Samuel Quinn). However, the RV data quality was insufficient to conclusively rule out the EB scenario. Ground-based photometry later retired TIC 188593930 as a false positive, because a deep eclipse was observed on a nearby faint star TIC 188593929~\footnote{\url{https://exofop.ipac.caltech.edu/}}. Therefore, we find that none of the light curves in our sample exhibit light curves indicative of a Jovian transiting planet.

\section{Planet Injection Pipeline} \label{s:injection}

To accurately measure the planet occurrence
rate based on the results of our planet search, we need
to determine the probability of detecting a planet using our pipeline.
We measure the detection efficiency of our planet detection pipeline by injecting transiting planet signals into the TESS light curves and running
the light curves through our detection pipeline where they will be detrended and vetted.
We generate synthetic transit signals for each light curve with orbital parameters drawn from a uniform distribution using pylightcurve~\citep{pylight}. Each synthetic planet signal that is injected into a light curve has a randomly determined planetary radius, orbital period, mid-time, and inclination (Rp, P, t$_0$, i) taken from the following uniform distributions:

$$\frac{P}{day} \sim U(0.5, 20)$$
$$\frac{R_p}{R_J} \sim U(0.08, 2)$$
$$t_0 \sim U(0, P)$$
$$i \sim U(i_{min}, i_{max})$$
\\
where $i_{max}$ and $i_{min}$ are calculated for each star as the maximum and minimum inclination for a transit using the semi-major axis and the radius each star. The eccentricity is set to 0, given that HJs are subject to strong orbital circularization and are expected to typically have low eccentricities~\citep[e.g., ][]{Montes2019}. Using stellar parameters, the limb darkening coefficients are determined by interpolating the \cite{Claret2017} limb darkening tables.  Specifically, they are estimated with the effective temperature, surface gravity, and the median metallicity of our sample (i.e. [Fe/H]=-1.27), which obtained from the TESS Input Catalog \citep{Stassun2018}.  After each light curve is modified to contain a synthetic planet signal, it is then sent through our planet detection pipeline. 

 \begin{figure}[t]
\begin{center}
\includegraphics[width=\linewidth]{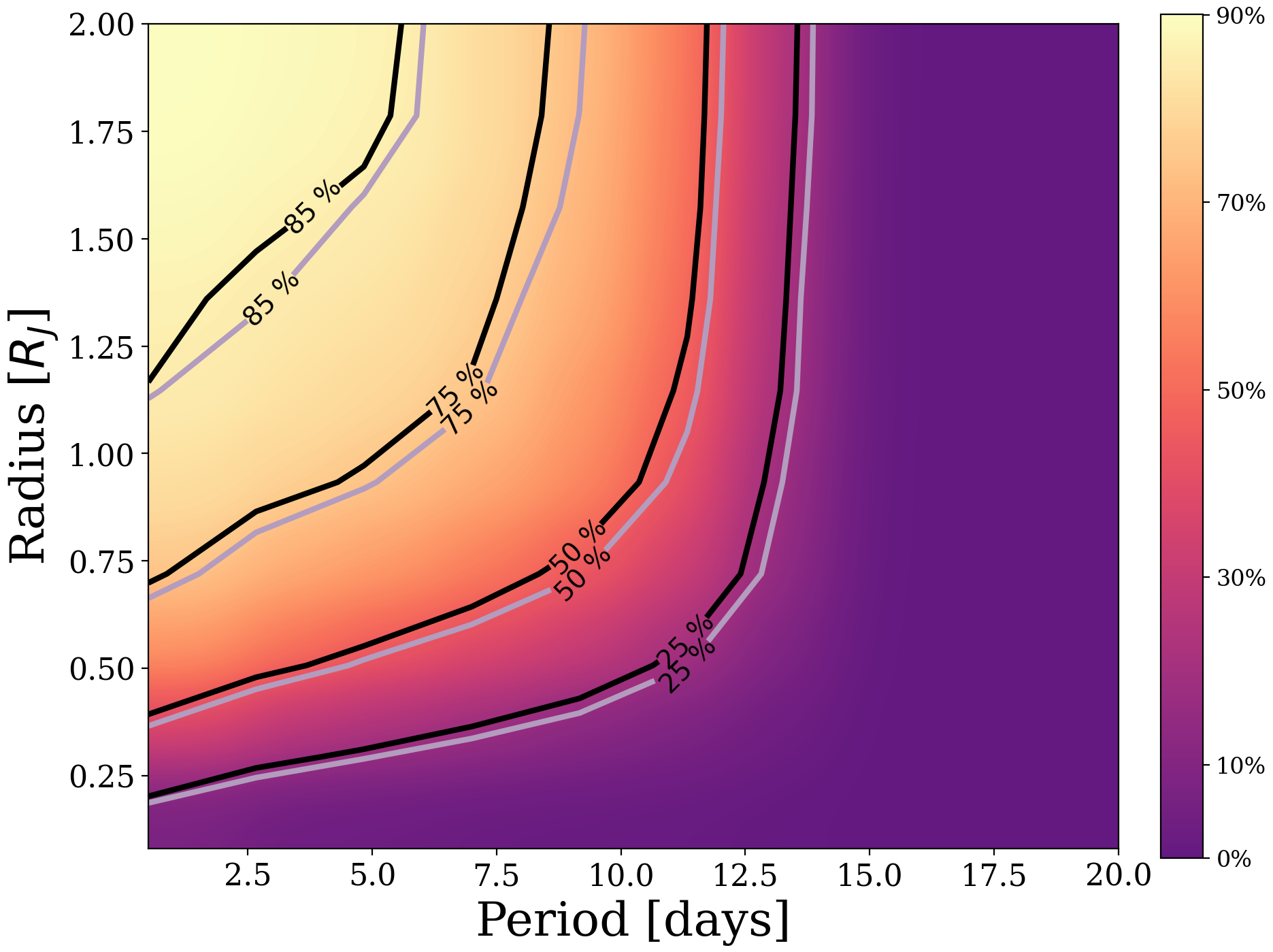}
\caption{Smoothed map comparing the detection efficiency of TLS (Gray) to BLS (Black). As indicated in the color bar which corresponds to the the detection efficiency of BLS, darker regions indicate lower detection efficiencies. The gray box indicates the region 0.8-2 $R_p$ and 0.5-10  days where BLS performs worse than TLS. \label{fig:tlsbls}}
\end{center}
\end{figure}

For a synthetic planet signal to be considered a detection, we used a nearly identical vetting process to the detection pipeline. We require the SNR to be greater than 5$\sigma$. The modified light curves must be determined by \texttt{Edivetter Unplugged} not to be a false positive. We also require the FAP value to be $\leq$ 0.0001. After the modified light curve undergoes this vetting process, the remaining synthetic planet signals are considered a detection if the detected period of the planet matches within 1$\%$ of the generated period. Period aliases of the injected signal are not treated as detections. It is important to note that, unlike real data, we do not conduct visual inspections as it would be highly impractical given the large number of injections. 

\subsection{Assessing our Pipeline}

To assess the detection efficiency of our pipeline, we conducted extensive injection-recovery tests. For each star, we generated 200 synthetic transit signals per light curve. In total, we injected 2,225,000 transiting planets into the light curves of the 11,125 stars for both TLS and BLS. Our pipeline successfully recovered $\sim$80$\%$ of injected planet signals with an SNR $\geq$ 5 and FAP $\leq$ 0.0001 within the range of $0.8-2$ $R_J$ and periods $0.5-10$ days with both TLS and BLS. We used both TLS and BLS and compare their results; however, we will focus on TLS within this section and elaborate on the difference between TLS and BLS within \S \ref{s:tvb}. 

We calculate the detection efficiency by dividing the number of the synthetic planet signals detected by the total planet signals injected. We take this fraction as the detection efficiency, which is binned as a function of planet period and radius. As shown in Figure \ref{fig:tlsbls}
, we found that our pipeline is sensitive to injected planets with radii larger than
0.8 $R_J$. Most of those planet signals were detected with nearly 75$\%$ efficiency out to the orbital period of
$\sim$ 10 days. Our pipeline displayed a significant decrease in the detection efficiency of planets with
radii $<$ 0.5 $R_J$  with the recovery fraction $\sim 19\%$  and 0.5-2.0 $R_J$ planets with periods longer than 10 days with the recovery fraction $\sim$12$\%$; however, the decrease in detection efficiency for smaller planets and longer periods does not affect our ability to constrain the occurrence rate for HJs. While we focus HJs in this work, we note that it is possible to detect close-in Neptune-sized planets given our detection efficiency.

\subsection{Comparison of TLS vs BLS} \label{s:tvb}

Although the BLS algorithm has become a standard tool for transit searches,  \cite{tls2009} developed a new planet transit search algorithm, TLS, which is optimized to detect small planets. Instead of a box-shaped template, they use a template based on actual transiting planets. Given our large data set, we determined that it would be pragmatic to compare TLS and BLS for HJs. 

As stated in Section \ref{s:detection}, our detection pipeline allows us to run both TLS and BLS. During our planet injection-recovery tests, we save the modified light curves, which are then sent through both TLS and BLS ensuring an impartial comparison. For TLS, we use the default template along with the limb darkening coefficients, mass, and radius for each star.  

We find that these algorithms are comparable; however, TLS slightly outperforms BLS. In Figure \ref{fig:tlsbls}, we show the detection efficiency. We find that overall TLS successfully recovered 38.8$\%$ of the injected planets using the default template, while BLS successfully recovered 37.4$\%$. However, TLS and BLS perform significantly better within the region 0.8-2 $R_J$ and 0.5-10 days that we consider for HJs. For HJs, TLS recovers 80.9$\%$ of the injected planets, whereas BLS recovers 79.5$\%$.

\section{Calculating Search Completeness} \label{s:completeness}

We define the search completeness to be the probability of finding a
planet with a random orbital alignment within our sample using our detection pipeline. Therefore, the overall search completeness depends both on the detectability, $P_{det}(R_p,P)$, of a particular planet and the likelihood that it will be observed
to transit. We determine the detection efficiency of our pipeline by conducting an extensive injection-recovery tests using our planet injection pipeline (see Section \ref{s:injection}).

Once the detection efficiency was determined, we use the geometric transit probability to account for the likelihood of a planet being oriented so that a transit can be observed. For the geometric transit probability we assume an eccentricity of zero and use: 

\begin{equation}\label{eq:transprob}
P_t(R_p,P)= \frac{R_p + R_\star}{a}
\end{equation}
where $R_\star$ is the star radius and \textit{a} the orbital semi-major
axis. \textit{a} is determined from Kepler's third law assuming a stellar mass of 1.01$M_\odot$ and radius of 0.711$R_\odot$, which are the median values for the halo stars within our sample. 

\begin{figure}[t]
\begin{center}
\includegraphics[width=\linewidth]{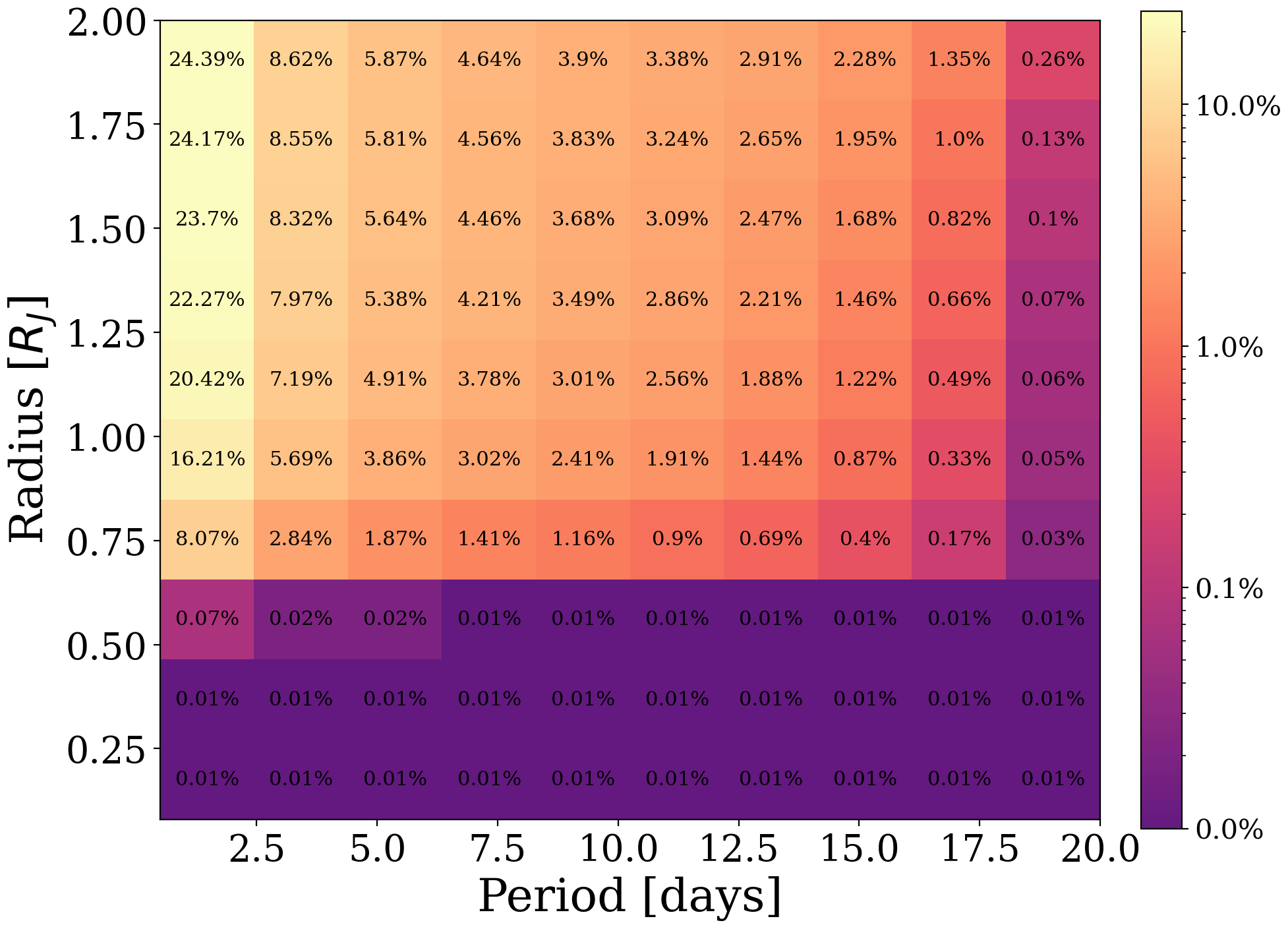}
\caption{The search completeness using BLS, showing the probability of observing a planet with a random orbital alignment within our sample using our detection pipeline. \label{fig:Completeness}}
\end{center}
\end{figure}

For a given orbital period and planet radius, we computed the corresponding semi-major axis for a planet orbiting each of the stars in our sample. With the transit probability for each radius and orbital period, we calculated the search completeness by weighting the detection efficiency by the geometric transit probability for each bin. In Figure \ref{fig:Completeness}, we show the search completeness.

\begin{figure*}[t]
\begin{center}
\includegraphics[width=0.85\linewidth]{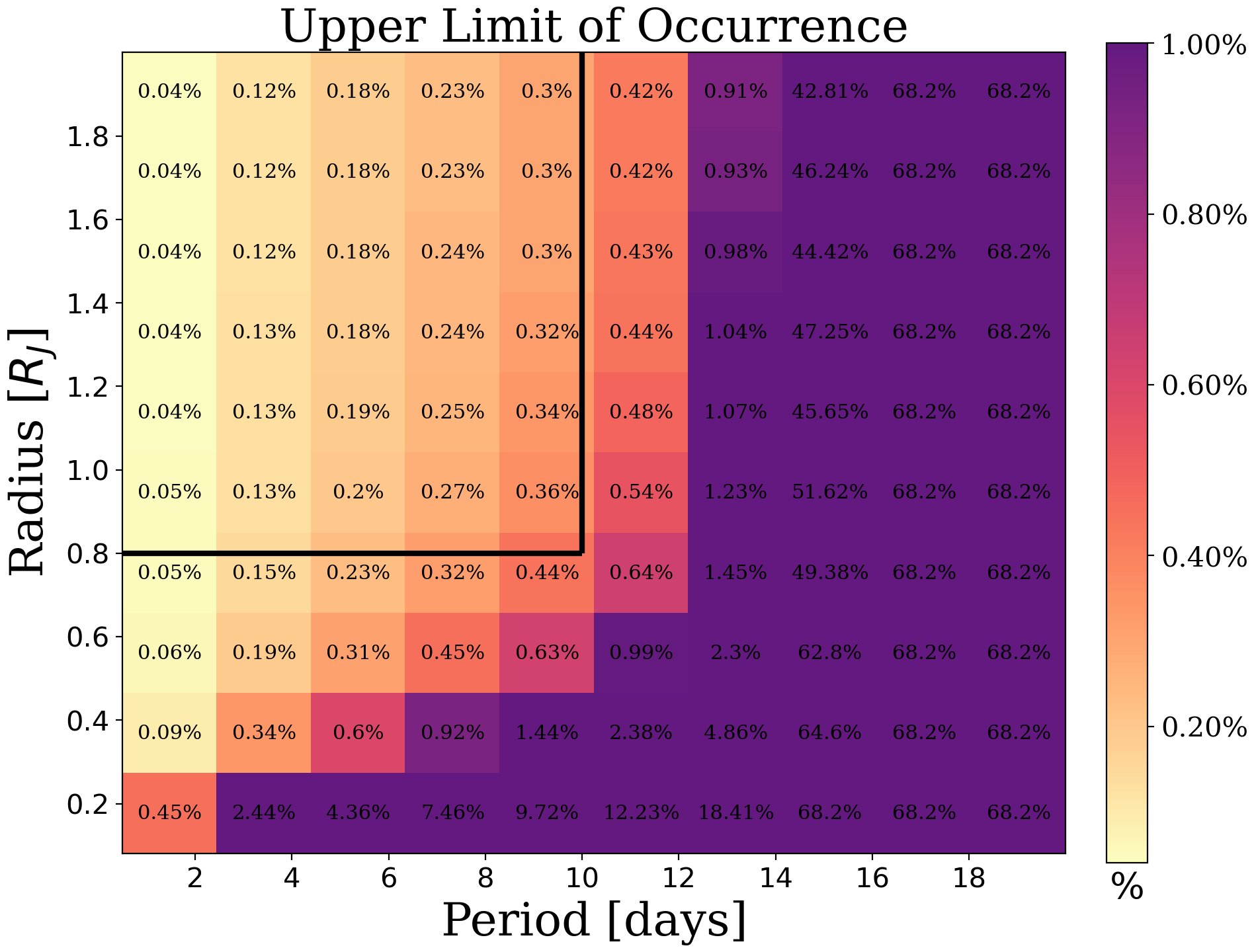}
\caption{ The region within the black box contains the binned upper limit of HJ occurrence within the metal-poor regime at the 1-$\sigma$ confidence interval. The numbers within each grid cell indicate the upper limit of HJ occurrence rate as a percentage. We assume stellar mass of 1.01$M_\odot$ and radius of 0.711$R_\odot$. \label{fig:occur}}
\end{center}
\end{figure*}

\section{Planet Occurrence Rate} \label{s:occur}
To determine the upper limit of planet occurrence given a null detection, we follow the procedure originally  described in the appendix of \cite{Burgasser2003} and in \cite{McCarthy2004} using the binomial distribution. The probability $P(f_p)$ of finding \textit{d} detections in a sample of size \textit{N} can be calculated as a function of the true planet frequency $f_p$:
\begin{equation} \label{eq:binomial}
P(d, N, f_p)= f_p^d  \hspace{1.5mm}(1-f_p)^{N-d} \hspace{1.5mm} \frac{N!}{(N-d)! \hspace{1mm}d\hspace{.5mm}!}
\end{equation}
\\
In the case of a null detection \textit{d} = 0, the planet frequency is constrained to the interval ($0,f_{p,max}$) for a confidence interval, C by 
\begin{equation} \label{eq:int}
\int_{0}^{f_{p,max}} (N+1) \times P(0, N, f_p) \hspace{1.5mm} df_p= C
\end{equation}
where $(N+1)$ is a normalization factor. Using this equation, the maximum planet frequency can be solved for

\begin{equation} \label{eq:freq}
f_{p,max}= 1- ( 1- C)^{\frac{1}{N+1}}
\end{equation}
\\
Therefore, we can calculate an upper limit of $f_p$ at a given confidence level by setting $C$ to the corresponding probability for the confidence level of interest. 

However, not all planets are detectable using the transit method and \textit{N} must be replaced by the effective sample size $N_{eff}$. To calculate the true upper limit of planet occurrence, we must account for the geometric transit probability and the detection efficiency of our pipeline. Therefore, we define our effective sample size as:

\begin{equation} \label{eq:N_eff}
N_{eff}(R_p, P)= N \times P_{t}(R_p,P) \times P_{det}(R_p,P)
\end{equation}
\\
where \textit{N} is the total sample size, $P_t(R_p,P)$ is the geometric transit probability, $P_{det}(R_p,P)$ is the detection efficiency (see Section \ref{s:completeness}), \textit{P} the orbital period and $R_p$ the planet radius.

Using equation(\ref{eq:freq}) and (\ref{eq:N_eff}) , we calculate the the upper limit of HJ occurrence by measuring the range in $f_p$ that covers 68$\%$ of the integrated probability function.This is equivalent to the 1-$\sigma$ confidence level. We conservatively choose this confidence level as to not overestimate the upper limit of HJ occurrence.

 \startlongtable
\begin{deluxetable*}{ccccccc}
\tablecaption{Hot Jupiter Occurrence Comparison\label{tab:1}}
\tablecolumns{7}
\tablehead{
\colhead{Author} & \colhead{N} & \colhead{$f_p$} & \colhead{[Fe/H]} & \colhead{P} &  \colhead{$R_p$} &  \colhead{Method}
}
\startdata
\cite{Petigura2018} & 1883$^\dagger$& $0.57^{+0.14}_{-0.12}  \%$ & (-0.3$\leq$[Fe/H]$\leq$ 0.3) & $1 \leq P\leq10$ days & 0.71-2.14$R_J$& Transit\\
\cite{Wright2012} & 836& $1.2\pm 0.38 \%$ & (-0.4$\leq$ [Fe/H] $\leq$ 0.3) & $\leq 10$ days & $>$0.51$R_J$& RV\\
\cite{Zhou2020}  &2401$^\dagger$& $0.4 \pm 0.10 \%$ & (-0.5$\leq$[Fe/H]$\leq$ 0.4) & $0.9 \leq P\leq10$ days & 0.8-2.5$R_J$ & Transit\\
\cite{Mayor2011} & 822 & $ 0.89 \pm 0.36\%$& (-0.5$\leq$ [Fe/H] $\leq$ 0.3) & $\leq 11$ days & $>$0.75 $R_J$ & RV\\
\cite{Masuda2017} & 1737$^\dagger$& $0.43^{+0.07}_{-0.06}  \%$ & (-0.6 $\leq$ [Fe/H] $\leq$ 0.2) & $0.5\leq P\leq10$ days & 0.8-2$R_J$& Transit\\
\cite{mortier2012} & 114 & 1.00$\%$$^\ast$& (-1.4$\leq$ [Fe/H] $\leq$ -0.4) & $\leq 10$ days & $>$1.04 $R_J$ & RV \\
\cite{sozzetti2009} & 160 & 0.67$\%$$^\ast$ &(-1.6$\leq$[Fe/H]$\leq$ -0.6) & $<3$ yrs & $>$1.25 $R_J$  & RV\\
This Work & 576$^\dagger$& 0.18$\%^\ast$ & (-2.0$\leq$[Fe/H]$\leq$ -0.6) & $0.5 \leq P\leq10$ days & 0.8-2$R_J$ & Transit\\
\enddata
\tablecomments{ This table is arranged in terms of the minimum metallicity. We use the 5$\%$ to 95$\%$ quantiles for each study. For all radial velocity surveys, we use the \cite{Chen2017} relation to convert from mass to radius.\\
\\
$^\dagger$ We estimate the median effective sample size using our detection efficiency\\
$^\ast$ This is the maximum 1-$\sigma$ upper limit of occurrence}

\end{deluxetable*}

In Figure \ref{fig:occur}, we show the our results using the detection efficiency determined from BLS. For the upper limit of HJ occurrence, we consider a radius of $0.8-2$ $R_J$ and periods $0.5-10$ days. We choose a radius of 2 $R_J$ given that the majority of the confirmed planets to date are less than 2 $R_J$ \citep{Exoarchive2013}. Within this region, we find upper limit of occurrence range between 0.04$\% \leq f_p\leq0.36\%$ within the parameter space considered with the mean upper limit of occurrence being $f_p<0.18\%$.

\section{Comparison to previous works} \label{s:comparison}

The occurrence of HJs has been studied by multiple surveys. For HJs orbiting solar-type stars, RV surveys have found the occurrence to be 
between $0.9 \% - 1.2 \%$ \citep{Wright2012,Marcy2005,Mayor2011,Petigura2018}, which is in contrast to the significantly lower occurrence rates determined with transit surveys. The HJ occurrence observed using TESS and Kepler range between $0.4\% - 0.6\%$ \citep{Howard2012,Fressin2013,Mulders2015,Santerne,Petigura2018, Zhou2020}. Although TESS and Kepler observe different targets, the occurrence rates are consistent. 

The discrepancy of occurrence rates between RV and transit surveys has been address by several works. Many studies have speculated that the difference may be a result of selection bias. \citet{Guo2017} investigated the differences in metallicity between Kepler and some radial velocity surveys; however, they found the difference in metallicity to be insufficient to explain the HJ occurrence discrepancy. Another possible explanation for the discrepancy may be attributed to the contamination of multiple stellar systems and evolved stars in transit samples~\citep{Wang2015}. Most recently, \cite{Maxwell2019} concluded that selection effects are the most probable cause of the HJ occurrence discrepancy. They quantitatively demonstrate that RV surveys
for giant planets can increase the occurrence rate by a factor of two by systematically removing spectroscopic binaries
from their samples.

The first attempt to determine HJ occurrence within the metal-poor regime was \cite{sozzetti2009}. They used radial velocity data  with most stars having 4 to 10 measurements from HIRES on the Keck 1 telescope with a sample of 160 metal-poor stars (see Table \ref{tab:1}). Consequently, strong constraints were derived for short-period gas giants where they had 95$\%$ completeness for periods $\leq 3$ yrs with a minimum companion mass of $\sim 0.75$ $M_J$, which corresponds to a HJ radius of approximately  $1.25$ $R_J$ \citep{Chen2017}. Given that the majority of the range is for cool Jupiters, we consider radius as a step function using the \cite{Chen2017} for periods $<10$ days and \cite{Thorngren2019} for periods $>10$ days.  Within the HJ range, we expect the planets to be inflated thereby increasing the radius. Their work resulted in a 1-$\sigma$ upper limit of HJ occurrence of 0.67$\%$.

The most recent work to directly consider this regime for HJs is \cite{mortier2012}. Similar to \cite{sozzetti2009}, radial velocity measurements were taken using HIRES and HARPS with a sample of 114 stars. To avoid biases, the sample was restricted to targets that had six or more measurements. By imposing this requirement, only 50 of the targets from \cite{sozzetti2009} were used in their work. For a completeness of 80$\%$ and period $<10$ days, they found a minimum companion mass of $\sim 0.314$ $M_J$, which corresponds to a radius of $1.04$ $R_J$ \citep{Chen2017}. Once again, we use the Mass- Radius relationship from \cite{Chen2017}, because we expect the HJs to be inflated. Given the smaller sample size, they found the 1-$\sigma$ upper limit of HJ occurrence to be 1.00$\%$.

Compared to the previous works using radial velocities, we place the most stringent upper limit on HJ occurrence. Using TESS FFI data, we allow for a larger effective sample size that ranges from $N_{eff} =313-2,582$ per bin with a median of $N_{eff}= 576$ (see equation \ref{eq:N_eff}). Our minimum effect sample size is $\sim 2-3$ times larger than the sample sizes of \cite{mortier2012} and \cite{sozzetti2009} (see Table \ref{tab:1}). Another difference to note is that each study has varying ranges of 
radius and period (see Table \ref{tab:1}); however, this does not significantly affect the comparison. For example, our upper limit of occurrence within the range \cite{mortier2012} considered is between  0.04$\% \leq f_p\leq 0.34\%$ with the mean upper limit of 0.17$\%$.

 From the results of this work, a metallicity limit can be established below which no HJs can form. Within \cite{mortier2012}, the authors suggested that giant planets can not form below $[Fe/H] = -0.5$. Similarly, we propose that the metallicity limit for HJs would be between -0.7 and -0.6, which is the upper limit of the metallicity range for this work. This value is consistent with observations and previous works \citep{Mordasini2012,mortier2012}. To date, no HJ has been discovered below a metallicity of -0.6 \citep{Exoarchive2013}. Currently, the most metal-poor star to host a HJ has a stellar metallicity of -0.6 \citep{Hellier2014}. 

 Although there is a strong correlation between metallicity, it is important to note that we are referring to the iron abundance of the star. However, there is evidence that indicates the overall metallicity of iron-poor HJ hosts is higher than the iron abundance \citep{Haywood2008,Adibekyan2012, Adibekyan2012a}. Stars of the thin and thick disks have an increase in $\alpha$ elements as a function of metallicity \citep{Haywood2008,Adibekyan2012a}. As a result, almost all giant planet hosts in the metal-poor regime have high [$\alpha$/Fe] values, such as Mg and Si \citep{Adibekyan2012a}. This indicates that overall metallicity is, in fact, higher than the iron-abundance and that $\alpha$ elements aid in the formation of HJs in the metal-poor regime.  
\\
\\
 
\section{Summary \& Conclusions} \label{s:summary and conclusions}

In this paper, we presented an updated upper limit of
HJ occurrence within the metal-poor regime using TESS data. We developed our own planet detection pipeline to search for transiting planets within our light curves. We then characterized the completeness of our pipeline by injecting simulated transiting planets into the light curves and
attempted to recover them. Our search of the light
curves of 11,125 halo stars revealed no planet candidates. We then calculated the upper limit of occurrence for HJs around halo stars by using the null detection hypotheses accounting for the geometric transit probability and the detection efficiency of our pipeline. The primary conclusions of this work are:

\begin{itemize}
  \item In line with previous works, HJs are rare or non-existent in the metal-poor regime (-2.0 $\leq$ [Fe/H] $\leq$ -0.6). 

  \item We further constrain the upper limit HJ occurrence within the metal-poor regime using TESS data. We find the upper limit of HJ occurrence range from 0.04$ \% \leq f_p\leq 0.36\%$ with the mean occurrence being $f_p< 0.18\%$ for radii 0.8 -2 R$_J$ and periods $0.5- 10$ days at the 1-$\sigma$ confidence level, which places the most stringent constraint on HJ  occurrence.
   
  \item TLS is comparable to the BLS when searching for HJs around metal-poor stars. It performs slightly better than BLS overall.
\end{itemize}

Further constraining the occurrence of HJ occurrence would require significantly larger samples; however, any planet discoveries would be of great interest within the metal-poor regime. They would expand our knowledge on the environment in which the first generation of planets formed. 

Moving forward, our future work will build upon this work. Using the framework that was developed conducting this study, we plan to determine an accurate functional form of planet occurrence vs. metallicity \citep{Johnson2010,Fischer2005,Udry2007}. We will consider a larger sample size and utilize the pipelines created for this work. This is a starting point in constructing a galactic planet formation model, which relates the metal-enrichment of the Milky Way to planet formation.  

\acknowledgments

 We thank David Latham for his effort in organizing the TESS Follow-up Observing Program Working Group. We acknowledge the MEarth team for the providing the follow-up observation.TSL is supported by NASA through Hubble Fellowship grant HST-HF2-51439.001, awarded by the Space Telescope Science Institute, which is operated by the Association of Universities for Research in Astronomy, Inc., for NASA, under contract NAS5-26555. JCZ is supported by an NSF Astronomy and Astrophysics Postdoctoral Fellowship under award AST-2001869. We also thank the anonymous referee and Johanna Teske who both provided suggestions that improved the quality and credibility of the manuscript.
 \software{ \texttt{eleanor}~\citep{2019PASP..131i4502F}, \texttt{pylightcurve}~\citep{pylight},  \texttt{transit least quares algorithm}~ \citep{tls2009}, and \texttt{astropy} ~ \citep{Astropy2013}}

\bibliography{Bib}

\begin{thebibliography}{}
\expandafter\ifx\csname natexlab\endcsname\relax\def\natexlab#1{#1}\fi
\providecommand{\url}[1]{\href{#1}{#1}}
\providecommand{\dodoi}[1]{doi:~\href{http://doi.org/#1}{\nolinkurl{#1}}}
\providecommand{\doeprint}[1]{\href{http://ascl.net/#1}{\nolinkurl{http://ascl.net/#1}}}
\providecommand{\doarXiv}[1]{\href{https://arxiv.org/abs/#1}{\nolinkurl{https://arxiv.org/abs/#1}}}

\bibitem[{{Adibekyan} {et~al.}(2012{\natexlab{a}}){Adibekyan}, {Delgado Mena},
  {Sousa}, {Santos}, {Israelian}, {Gonz{\'a}lez Hern{\'a}ndez}, {Mayor}, \&
  {Hakobyan}}]{Adibekyan2012}
{Adibekyan}, V.~Z., {Delgado Mena}, E., {Sousa}, S.~G., {et~al.}
  2012{\natexlab{a}}, \aap, 547, A36, \dodoi{10.1051/0004-6361/201220167}

\bibitem[{{Adibekyan} {et~al.}(2012{\natexlab{b}}){Adibekyan}, {Santos},
  {Sousa}, {Israelian}, {Delgado Mena}, {Gonz{\'a}lez Hern{\'a}ndez}, {Mayor},
  {Lovis}, \& {Udry}}]{Adibekyan2012a}
{Adibekyan}, V.~Z., {Santos}, N.~C., {Sousa}, S.~G., {et~al.}
  2012{\natexlab{b}}, \aap, 543, A89, \dodoi{10.1051/0004-6361/201219564}

\bibitem[{{Akeson} {et~al.}(2013){Akeson}, {Chen}, {Ciardi}, {Crane}, {Good},
  {Harbut}, {Jackson}, {Kane}, {Laity}, {Leifer}, {Lynn}, {McElroy}, {Papin},
  {Plavchan}, {Ram{\'\i}rez}, {Rey}, {von Braun}, {Wittman}, {Abajian}, {Ali},
  {Beichman}, {Beekley}, {Berriman}, {Berukoff}, {Bryden}, {Chan}, {Groom},
  {Lau}, {Payne}, {Regelson}, {Saucedo}, {Schmitz}, {Stauffer}, {Wyatt}, \&
  {Zhang}}]{Exoarchive2013}
{Akeson}, R.~L., {Chen}, X., {Ciardi}, D., {et~al.} 2013, \pasp, 125, 989,
  \dodoi{10.1086/672273}

\bibitem[{{Alvarado-Montes} \& {Garc{\'\i}a-Carmona}(2019)}]{Montes2019}
{Alvarado-Montes}, J.~A., \& {Garc{\'\i}a-Carmona}, C. 2019, \mnras, 486, 3963,
  \dodoi{10.1093/mnras/stz1081}

\bibitem[{{Arriagada}(2011)}]{Arriagada2011}
{Arriagada}, P. 2011, \apj, 734, 70, \dodoi{10.1088/0004-637X/734/1/70}

\bibitem[{{Astropy Collaboration} {et~al.}(2013){Astropy Collaboration},
  {Robitaille}, {Tollerud}, {Greenfield}, {Droettboom}, {Bray}, {Aldcroft},
  {Davis}, {Ginsburg}, {Price-Whelan}, {Kerzendorf}, {Conley}, {Crighton},
  {Barbary}, {Muna}, {Ferguson}, {Grollier}, {Parikh}, {Nair}, {Unther},
  {Deil}, {Woillez}, {Conseil}, {Kramer}, {Turner}, {Singer}, {Fox}, {Weaver},
  {Zabalza}, {Edwards}, {Azalee Bostroem}, {Burke}, {Casey}, {Crawford},
  {Dencheva}, {Ely}, {Jenness}, {Labrie}, {Lim}, {Pierfederici}, {Pontzen},
  {Ptak}, {Refsdal}, {Servillat}, \& {Streicher}}]{Astropy2013}
{Astropy Collaboration}, {Robitaille}, T.~P., {Tollerud}, E.~J., {et~al.} 2013,
  \aap, 558, A33, \dodoi{10.1051/0004-6361/201322068}

\bibitem[{{Bailey} \& {Batygin}(2018)}]{Bailey2018}
{Bailey}, E., \& {Batygin}, K. 2018, \apjl, 866, L2,
  \dodoi{10.3847/2041-8213/aade90}

\bibitem[{{Bernstein} {et~al.}(2003){Bernstein}, {Shectman}, {Gunnels},
  {Mochnacki}, \& {Athey}}]{Bernstein03}
{Bernstein}, R., {Shectman}, S.~A., {Gunnels}, S.~M., {Mochnacki}, S., \&
  {Athey}, A.~E. 2003, Proc. SPIE, 4841, 1694, \dodoi{10.1117/12.461502}

\bibitem[{{Burgasser} {et~al.}(2003){Burgasser}, {Kirkpatrick}, {Reid},
  {Brown}, {Miskey}, \& {Gizis}}]{Burgasser2003}
{Burgasser}, A.~J., {Kirkpatrick}, J.~D., {Reid}, I.~N., {et~al.} 2003, \apj,
  586, 512, \dodoi{10.1086/346263}

\bibitem[{{Chen} \& {Kipping}(2017)}]{Chen2017}
{Chen}, J., \& {Kipping}, D. 2017, \apj, 834, 17,
  \dodoi{10.3847/1538-4357/834/1/17}

\bibitem[{{Choi} \& {Nagamine}(2009)}]{Choi2009}
{Choi}, J.-H., \& {Nagamine}, K. 2009, \mnras, 393, 1595,
  \dodoi{10.1111/j.1365-2966.2008.14297.x}

\bibitem[{{Claret}(2017)}]{Claret2017}
{Claret}, A. 2017, \aap, 600, A30, \dodoi{10.1051/0004-6361/201629705}

\bibitem[{{Claret} \& {Bloemen}(2011)}]{claret}
{Claret}, A., \& {Bloemen}, S. 2011, \aap, 529, A75,
  \dodoi{10.1051/0004-6361/201116451}

\bibitem[{{Cumming} {et~al.}(1999){Cumming}, {Marcy}, \&
  {Butler}}]{Cumming1999}
{Cumming}, A., {Marcy}, G.~W., \& {Butler}, R.~P. 1999, \apj, 526, 890,
  \dodoi{10.1086/308020}

\bibitem[{Dressing \& Charbonneau(2015)}]{Dressing2015}
Dressing, C.~D., \& Charbonneau, D. 2015, The Astrophysical Journal, 807, 45,
  \dodoi{10.1088/0004-637x/807/1/45}

\bibitem[{{Endl} {et~al.}(2006{\natexlab{a}}){Endl}, {Cochran}, {K{\"u}rster},
  {Paulson}, {Wittenmyer}, {MacQueen}, \& {Tull}}]{2006ApJ...649..436E}
{Endl}, M., {Cochran}, W.~D., {K{\"u}rster}, M., {et~al.} 2006{\natexlab{a}},
  \apj, 649, 436, \dodoi{10.1086/506465}

\bibitem[{{Endl} {et~al.}(2006{\natexlab{b}}){Endl}, {Cochran}, {K{\"u}rster},
  {Paulson}, {Wittenmyer}, {MacQueen}, \& {Tull}}]{Endl2006}
---. 2006{\natexlab{b}}, \apj, 649, 436, \dodoi{10.1086/506465}

\bibitem[{{Ercolano} \& {Clarke}(2010)}]{Ercolano2010}
{Ercolano}, B., \& {Clarke}, C.~J. 2010, \mnras, 402, 2735,
  \dodoi{10.1111/j.1365-2966.2009.16094.x}

\bibitem[{{Feinstein} {et~al.}(2019){Feinstein}, {Montet}, {Foreman-Mackey},
  {Bedell}, {Saunders}, {Bean}, {Christiansen}, {Hedges}, {Luger}, {Scolnic},
  \& {Cardoso}}]{2019PASP..131i4502F}
{Feinstein}, A.~D., {Montet}, B.~T., {Foreman-Mackey}, D., {et~al.} 2019,
  \pasp, 131, 094502, \dodoi{10.1088/1538-3873/ab291c}

\bibitem[{{Fischer} \& {Valenti}(2005)}]{Fischer2005}
{Fischer}, D.~A., \& {Valenti}, J. 2005, \apj, 622, 1102,
  \dodoi{10.1086/428383}

\bibitem[{{Forgan} {et~al.}(2017){Forgan}, {Rowlands}, {Gomez}, {Gomez},
  {Schofield}, {Dunne}, \& {Maddox}}]{Forgan2017}
{Forgan}, D.~H., {Rowlands}, K., {Gomez}, H.~L., {et~al.} 2017, \mnras, 472,
  2289, \dodoi{10.1093/mnras/stx2162}

\bibitem[{{Fressin} {et~al.}(2013){Fressin}, {Torres}, {Charbonneau}, {Bryson},
  {Christiansen}, {Dressing}, {Jenkins}, {Walkowicz}, \&
  {Batalha}}]{Fressin2013}
{Fressin}, F., {Torres}, G., {Charbonneau}, D., {et~al.} 2013, \apj, 766, 81,
  \dodoi{10.1088/0004-637X/766/2/81}

\bibitem[{{Gaia Collaboration} {et~al.}(2018){Gaia Collaboration}, {Brown},
  {Vallenari}, {Prusti}, {de Bruijne}, {Babusiaux}, {Bailer-Jones}, {Biermann},
  {Evans}, {Eyer}, {Jansen}, {Jordi}, {Klioner}, {Lammers}, {Lindegren},
  {Luri}, {Mignard}, {Panem}, {Pourbaix}, {Randich}, {Sartoretti}, {Siddiqui},
  {Soubiran}, {van Leeuwen}, {Walton}, {Arenou}, {Bastian}, {Cropper},
  {Drimmel}, {Katz}, {Lattanzi}, {Bakker}, {Cacciari}, {Casta{\~n}eda},
  {Chaoul}, {Cheek}, {De Angeli}, {Fabricius}, {Guerra}, {Holl}, {Masana},
  {Messineo}, {Mowlavi}, {Nienartowicz}, {Panuzzo}, {Portell}, {Riello},
  {Seabroke}, {Tanga}, {Th{\'e}venin}, {Gracia-Abril}, {Comoretto},
  {Garcia-Reinaldos}, {Teyssier}, {Altmann}, {Andrae}, {Audard},
  {Bellas-Velidis}, {Benson}, {Berthier}, {Blomme}, {Burgess}, {Busso},
  {Carry}, {Cellino}, {Clementini}, {Clotet}, {Creevey}, {Davidson}, {De
  Ridder}, {Delchambre}, {Dell'Oro}, {Ducourant},
  {Fern{\'a}ndez-Hern{\'a}ndez}, {Fouesneau}, {Fr{\'e}mat}, {Galluccio},
  {Garc{\'\i}a-Torres}, {Gonz{\'a}lez-N{\'u}{\~n}ez}, {Gonz{\'a}lez-Vidal},
  {Gosset}, {Guy}, {Halbwachs}, {Hambly}, {Harrison}, {Hern{\'a}ndez},
  {Hestroffer}, {Hodgkin}, {Hutton}, {Jasniewicz}, {Jean-Antoine-Piccolo},
  {Jordan}, {Korn}, {Krone-Martins}, {Lanzafame}, {Lebzelter}, {L{\"o}ffler},
  {Manteiga}, {Marrese}, {Mart{\'\i}n-Fleitas}, {Moitinho}, {Mora}, {Muinonen},
  {Osinde}, {Pancino}, {Pauwels}, {Petit}, {Recio-Blanco}, {Richards},
  {Rimoldini}, {Robin}, {Sarro}, {Siopis}, {Smith}, {Sozzetti}, {S{\"u}veges},
  {Torra}, {van Reeven}, {Abbas}, {Abreu Aramburu}, {Accart}, {Aerts},
  {Altavilla}, {{\'A}lvarez}, {Alvarez}, {Alves}, {Anderson}, {Andrei},
  {Anglada Varela}, {Antiche}, {Antoja}, {Arcay}, {Astraatmadja}, {Bach},
  {Baker}, {Balaguer-N{\'u}{\~n}ez}, {Balm}, {Barache}, {Barata}, {Barbato},
  {Barblan}, {Barklem}, {Barrado}, {Barros}, {Barstow}, {Bartholom{\'e}
  Mu{\~n}oz}, {Bassilana}, {Becciani}, {Bellazzini}, {Berihuete}, {Bertone},
  {Bianchi}, {Bienaym{\'e}}, {Blanco-Cuaresma}, {Boch}, {Boeche}, {Bombrun},
  {Borrachero}, {Bossini}, {Bouquillon}, {Bourda}, {Bragaglia}, {Bramante},
  {Breddels}, {Bressan}, {Brouillet}, {Br{\"u}semeister}, {Brugaletta},
  {Bucciarelli}, {Burlacu}, {Busonero}, {Butkevich}, {Buzzi}, {Caffau},
  {Cancelliere}, {Cannizzaro}, {Cantat-Gaudin}, {Carballo}, {Carlucci},
  {Carrasco}, {Casamiquela}, {Castellani}, {Castro-Ginard}, {Charlot},
  {Chemin}, {Chiavassa}, {Cocozza}, {Costigan}, {Cowell}, {Crifo}, {Crosta},
  {Crowley}, {Cuypers}, {Dafonte}, {Damerdji}, {Dapergolas}, {David}, {David},
  {de Laverny}, {De Luise}, {De March}, {de Martino}, {de Souza}, {de Torres},
  {Debosscher}, {del Pozo}, {Delbo}, {Delgado}, {Delgado}, {Di Matteo},
  {Diakite}, {Diener}, {Distefano}, {Dolding}, {Drazinos}, {Dur{\'a}n},
  {Edvardsson}, {Enke}, {Eriksson}, {Esquej}, {Eynard Bontemps}, {Fabre},
  {Fabrizio}, {Faigler}, {Falc{\~a}o}, {Farr{\`a}s Casas}, {Federici},
  {Fedorets}, {Fernique}, {Figueras}, {Filippi}, {Findeisen}, {Fonti},
  {Fraile}, {Fraser}, {Fr{\'e}zouls}, {Gai}, {Galleti}, {Garabato},
  {Garc{\'\i}a-Sedano}, {Garofalo}, {Garralda}, {Gavel}, {Gavras}, {Gerssen},
  {Geyer}, {Giacobbe}, {Gilmore}, {Girona}, {Giuffrida}, {Glass}, {Gomes},
  {Granvik}, {Gueguen}, {Guerrier}, {Guiraud}, {Guti{\'e}rrez-S{\'a}nchez},
  {Haigron}, {Hatzidimitriou}, {Hauser}, {Haywood}, {Heiter}, {Helmi}, {Heu},
  {Hilger}, {Hobbs}, {Hofmann}, {Holland}, {Huckle}, {Hypki}, {Icardi},
  {Jan{\ss}en}, {Jevardat de Fombelle}, {Jonker}, {Juh{\'a}sz}, {Julbe},
  {Karampelas}, {Kewley}, {Klar}, {Kochoska}, {Kohley}, {Kolenberg},
  {Kontizas}, {Kontizas}, {Koposov}, {Kordopatis}, {Kostrzewa-Rutkowska},
  {Koubsky}, {Lambert}, {Lanza}, {Lasne}, {Lavigne}, {Le Fustec}, {Le
  Poncin-Lafitte}, {Lebreton}, {Leccia}, {Leclerc}, {Lecoeur-Taibi},
  {Lenhardt}, {Leroux}, {Liao}, {Licata}, {Lindstr{\o}m}, {Lister}, {Livanou},
  {Lobel}, {L{\'o}pez}, {Managau}, {Mann}, {Mantelet}, {Marchal}, {Marchant},
  {Marconi}, {Marinoni}, {Marschalk{\'o}}, {Marshall}, {Martino}, {Marton},
  {Mary}, {Massari}, {Matijevi{\v{c}}}, {Mazeh}, {McMillan}, {Messina},
  {Michalik}, {Millar}, {Molina}, {Molinaro}, {Moln{\'a}r}, {Montegriffo},
  {Mor}, {Morbidelli}, {Morel}, {Morris}, {Mulone}, {Muraveva}, {Musella},
  {Nelemans}, {Nicastro}, {Noval}, {O'Mullane}, {Ord{\'e}novic},
  {Ord{\'o}{\~n}ez-Blanco}, {Osborne}, {Pagani}, {Pagano}, {Pailler},
  {Palacin}, {Palaversa}, {Panahi}, {Pawlak}, {Piersimoni}, {Pineau}, {Plachy},
  {Plum}, {Poggio}, {Poujoulet}, {Pr{\v{s}}a}, {Pulone}, {Racero}, {Ragaini},
  {Rambaux}, {Ramos-Lerate}, {Regibo}, {Reyl{\'e}}, {Riclet}, {Ripepi}, {Riva},
  {Rivard}, {Rixon}, {Roegiers}, {Roelens}, {Romero-G{\'o}mez}, {Rowell},
  {Royer}, {Ruiz-Dern}, {Sadowski}, {Sagrist{\`a} Sell{\'e}s}, {Sahlmann},
  {Salgado}, {Salguero}, {Sanna}, {Santana-Ros}, {Sarasso}, {Savietto},
  {Schultheis}, {Sciacca}, {Segol}, {Segovia}, {S{\'e}gransan}, {Shih},
  {Siltala}, {Silva}, {Smart}, {Smith}, {Solano}, {Solitro}, {Sordo}, {Soria
  Nieto}, {Souchay}, {Spagna}, {Spoto}, {Stampa}, {Steele},
  {Steidelm{\"u}ller}, {Stephenson}, {Stoev}, {Suess}, {Surdej}, {Szabados},
  {Szegedi-Elek}, {Tapiador}, {Taris}, {Tauran}, {Taylor}, {Teixeira},
  {Terrett}, {Teyssand ier}, {Thuillot}, {Titarenko}, {Torra Clotet}, {Turon},
  {Ulla}, {Utrilla}, {Uzzi}, {Vaillant}, {Valentini}, {Valette}, {van Elteren},
  {Van Hemelryck}, {van Leeuwen}, {Vaschetto}, {Vecchiato}, {Veljanoski},
  {Viala}, {Vicente}, {Vogt}, {von Essen}, {Voss}, {Votruba}, {Voutsinas},
  {Walmsley}, {Weiler}, {Wertz}, {Wevers}, {Wyrzykowski}, {Yoldas},
  {{\v{Z}}erjal}, {Ziaeepour}, {Zorec}, {Zschocke}, {Zucker}, {Zurbach}, \&
  {Zwitter}}]{Gaia2018}
{Gaia Collaboration}, {Brown}, A.~G.~A., {Vallenari}, A., {et~al.} 2018, \aap,
  616, A1, \dodoi{10.1051/0004-6361/201833051}

\bibitem[{{Gonzalez} {et~al.}(2001){Gonzalez}, {Laws}, {Tyagi}, \&
  {Reddy}}]{gon1997}
{Gonzalez}, G., {Laws}, C., {Tyagi}, S., \& {Reddy}, B.~E. 2001, \aj, 121, 432,
  \dodoi{10.1086/318048}

\bibitem[{{Guo} {et~al.}(2017){Guo}, {Johnson}, {Mann}, {Kraus}, {Curtis}, \&
  {Latham}}]{Guo2017}
{Guo}, X., {Johnson}, J.~A., {Mann}, A.~W., {et~al.} 2017, \apj, 838, 25,
  \dodoi{10.3847/1538-4357/aa6004}

\bibitem[{{Hartman} \& {Bakos}(2016)}]{Vartools2016}
{Hartman}, J.~D., \& {Bakos}, G.~{\'A}. 2016, Astronomy and Computing, 17, 1,
  \dodoi{10.1016/j.ascom.2016.05.006}

\bibitem[{{Haworth} {et~al.}(2016){Haworth}, {Ilee}, {Forgan}, {Facchini},
  {Price}, {Boneberg}, {Booth}, {Clarke}, {Gonzalez}, {Hutchison}, {Kamp},
  {Laibe}, {Lyra}, {Meru}, {Mohanty}, {Pani{\'c}}, {Rice}, {Suzuki}, {Teague},
  {Walsh}, {Woitke}, \& {Community authors}}]{Haworth2016}
{Haworth}, T.~J., {Ilee}, J.~D., {Forgan}, D.~H., {et~al.} 2016, \pasa, 33,
  e053, \dodoi{10.1017/pasa.2016.45}

\bibitem[{{Haywood}(2008)}]{Haywood2008}
{Haywood}, M. 2008, \aap, 482, 673, \dodoi{10.1051/0004-6361:20079141}

\bibitem[{{Hellier} {et~al.}(2014){Hellier}, {Anderson}, {Collier Cameron},
  {Delrez}, {Gillon}, {Jehin}, {Lendl}, {Maxted}, {Pepe}, {Pollacco}, {Queloz},
  {S{\'e}gransan}, {Smalley}, {Smith}, {Southworth}, {Triaud}, {Udry}, \&
  {West}}]{Hellier2014}
{Hellier}, C., {Anderson}, D.~R., {Collier Cameron}, A., {et~al.} 2014, \mnras,
  440, 1982, \dodoi{10.1093/mnras/stu410}

\bibitem[{{Hippke} {et~al.}(2019){Hippke}, {David}, {Mulders}, \&
  {Heller}}]{wotan2019}
{Hippke}, M., {David}, T.~J., {Mulders}, G.~D., \& {Heller}, R. 2019, \aj, 158,
  143, \dodoi{10.3847/1538-3881/ab3984}

\bibitem[{{Hippke} \& {Heller}(2019)}]{tls2009}
{Hippke}, M., \& {Heller}, R. 2019, \aap, 623, A39,
  \dodoi{10.1051/0004-6361/201834672}

\bibitem[{{Howard} {et~al.}(2012){Howard}, {Marcy}, {Bryson}, {Jenkins},
  {Rowe}, {Batalha}, {Borucki}, {Koch}, {Dunham}, {Gautier}, {Van Cleve},
  {Cochran}, {Latham}, {Lissauer}, {Torres}, {Brown}, {Gilliland}, {Buchhave},
  {Caldwell}, {Christensen-Dalsgaard}, {Ciardi}, {Fressin}, {Haas}, {Howell},
  {Kjeldsen}, {Seager}, {Rogers}, {Sasselov}, {Steffen}, {Basri},
  {Charbonneau}, {Christiansen}, {Clarke}, {Dupree}, {Fabrycky}, {Fischer},
  {Ford}, {Fortney}, {Tarter}, {Girouard}, {Holman}, {Johnson}, {Klaus},
  {Machalek}, {Moorhead}, {Morehead}, {Ragozzine}, {Tenenbaum}, {Twicken},
  {Quinn}, {Isaacson}, {Shporer}, {Lucas}, {Walkowicz}, {Welsh}, {Boss},
  {Devore}, {Gould}, {Smith}, {Morris}, {Prsa}, {Morton}, {Still}, {Thompson},
  {Mullally}, {Endl}, \& {MacQueen}}]{Howard2012}
{Howard}, A.~W., {Marcy}, G.~W., {Bryson}, S.~T., {et~al.} 2012, \apjs, 201,
  15, \dodoi{10.1088/0067-0049/201/2/15}

\bibitem[{{Johnson} {et~al.}(2010){Johnson}, {Aller}, {Howard}, \&
  {Crepp}}]{Johnson2010}
{Johnson}, J.~A., {Aller}, K.~M., {Howard}, A.~W., \& {Crepp}, J.~R. 2010,
  \pasp, 122, 905, \dodoi{10.1086/655775}

\bibitem[{{Johnson} \& {Apps}(2009)}]{Johnson2009}
{Johnson}, J.~A., \& {Apps}, K. 2009, \apj, 699, 933,
  \dodoi{10.1088/0004-637X/699/2/933}

\bibitem[{Johnson \& Li(2012)}]{Johnson2012}
Johnson, J.~L., \& Li, H. 2012, The Astrophysical Journal, 751, 81,
  \dodoi{10.1088/0004-637x/751/2/81}

\bibitem[{{Koppelman} {et~al.}(2018){Koppelman}, {Helmi}, \&
  {Veljanoski}}]{Koppelman2018}
{Koppelman}, H., {Helmi}, A., \& {Veljanoski}, J. 2018, \apjl, 860, L11,
  \dodoi{10.3847/2041-8213/aac882}

\bibitem[{{Kornet} {et~al.}(2005){Kornet}, {Bodenheimer}, {R{\'o}{\.z}yczka},
  \& {Stepinski}}]{Kornet2005}
{Kornet}, K., {Bodenheimer}, P., {R{\'o}{\.z}yczka}, M., \& {Stepinski}, T.~F.
  2005, \aap, 430, 1133, \dodoi{10.1051/0004-6361:20041692}

\bibitem[{{Kov{\'a}cs} {et~al.}(2002){Kov{\'a}cs}, {Zucker}, \&
  {Mazeh}}]{Kovacs2002}
{Kov{\'a}cs}, G., {Zucker}, S., \& {Mazeh}, T. 2002, \aap, 391, 369,
  \dodoi{10.1051/0004-6361:20020802}

\bibitem[{{Marcy} {et~al.}(2005){Marcy}, {Butler}, {Fischer}, {Vogt}, {Wright},
  {Tinney}, \& {Jones}}]{Marcy2005}
{Marcy}, G., {Butler}, R.~P., {Fischer}, D., {et~al.} 2005, Progress of
  Theoretical Physics Supplement, 158, 24, \dodoi{10.1143/PTPS.158.24}

\bibitem[{{Masuda} \& {Winn}(2017)}]{Masuda2017}
{Masuda}, K., \& {Winn}, J.~N. 2017, \aj, 153, 187,
  \dodoi{10.3847/1538-3881/aa647c}

\bibitem[{{Mayor} {et~al.}(2011){Mayor}, {Marmier}, {Lovis}, {Udry},
  {S{\'e}gransan}, {Pepe}, {Benz}, {Bertaux}, {Bouchy}, {Dumusque}, {Lo Curto},
  {Mordasini}, {Queloz}, \& {Santos}}]{Mayor2011}
{Mayor}, M., {Marmier}, M., {Lovis}, C., {et~al.} 2011, arXiv e-prints,
  arXiv:1109.2497.
\newblock \doarXiv{1109.2497}

\bibitem[{{McCarthy} \& {Zuckerman}(2004)}]{McCarthy2004}
{McCarthy}, C., \& {Zuckerman}, B. 2004, \aj, 127, 2871, \dodoi{10.1086/383559}

\bibitem[{{Miller} \& {Fortney}(2011)}]{Miller2011}
{Miller}, N., \& {Fortney}, J.~J. 2011, \apjl, 736, L29,
  \dodoi{10.1088/2041-8205/736/2/L29}

\bibitem[{{Moe} \& {Kratter}(2019)}]{Maxwell2019}
{Moe}, M., \& {Kratter}, K.~M. 2019, arXiv e-prints, arXiv:1912.01699.
\newblock \doarXiv{1912.01699}

\bibitem[{{Mordasini} {et~al.}(2012){Mordasini}, {Alibert}, {Benz}, {Klahr}, \&
  {Henning}}]{Mordasini2012}
{Mordasini}, C., {Alibert}, Y., {Benz}, W., {Klahr}, H., \& {Henning}, T. 2012,
  \aap, 541, A97, \dodoi{10.1051/0004-6361/201117350}

\bibitem[{{Mortier} {et~al.}(2013){Mortier}, {Santos}, {Sousa}, {Israelian},
  {Mayor}, \& {Udry}}]{mortier13}
{Mortier}, A., {Santos}, N.~C., {Sousa}, S., {et~al.} 2013, \aap, 551, A112,
  \dodoi{10.1051/0004-6361/201220707}

\bibitem[{{Mortier} {et~al.}(2012){Mortier}, {Santos}, {Sozzetti}, {Mayor},
  {Latham}, {Bonfils}, \& {Udry}}]{mortier2012}
{Mortier}, A., {Santos}, N.~C., {Sozzetti}, A., {et~al.} 2012, \aap, 543, A45,
  \dodoi{10.1051/0004-6361/201118651}

\bibitem[{{Mulders} {et~al.}(2015){Mulders}, {Pascucci}, \&
  {Apai}}]{Mulders2015}
{Mulders}, G.~D., {Pascucci}, I., \& {Apai}, D. 2015, \apj, 798, 112,
  \dodoi{10.1088/0004-637X/798/2/112}

\bibitem[{{Petigura} {et~al.}(2018){Petigura}, {Marcy}, {Winn}, {Weiss},
  {Fulton}, {Howard}, {Sinukoff}, {Isaacson}, {Morton}, \&
  {Johnson}}]{Petigura2018}
{Petigura}, E.~A., {Marcy}, G.~W., {Winn}, J.~N., {et~al.} 2018, \aj, 155, 89,
  \dodoi{10.3847/1538-3881/aaa54c}

\bibitem[{{Pont} {et~al.}(2006){Pont}, {Zucker}, \& {Queloz}}]{Pont2006}
{Pont}, F., {Zucker}, S., \& {Queloz}, D. 2006, \mnras, 373, 231,
  \dodoi{10.1111/j.1365-2966.2006.11012.x}

\bibitem[{{Ricker} {et~al.}(2015){Ricker}, {Winn}, {Vanderspek}, {Latham},
  {Bakos}, {Bean}, {Berta-Thompson}, {Brown}, {Buchhave}, {Butler}, {Butler},
  {Chaplin}, {Charbonneau}, {Christensen-Dalsgaard}, {Clampin}, {Deming},
  {Doty}, {De Lee}, {Dressing}, {Dunham}, {Endl}, {Fressin}, {Ge}, {Henning},
  {Holman}, {Howard}, {Ida}, {Jenkins}, {Jernigan}, {Johnson}, {Kaltenegger},
  {Kawai}, {Kjeldsen}, {Laughlin}, {Levine}, {Lin}, {Lissauer}, {MacQueen},
  {Marcy}, {McCullough}, {Morton}, {Narita}, {Paegert}, {Palle}, {Pepe},
  {Pepper}, {Quirrenbach}, {Rinehart}, {Sasselov}, {Sato}, {Seager},
  {Sozzetti}, {Stassun}, {Sullivan}, {Szentgyorgyi}, {Torres}, {Udry}, \&
  {Villasenor}}]{TESS}
{Ricker}, G.~R., {Winn}, J.~N., {Vanderspek}, R., {et~al.} 2015, Journal of
  Astronomical Telescopes, Instruments, and Systems, 1, 014003,
  \dodoi{10.1117/1.JATIS.1.1.014003}

\bibitem[{{Santerne} {et~al.}(2016){Santerne}, {Moutou}, {Tsantaki}, {Bouchy},
  {H{\'e}brard}, {Adibekyan}, {Almenara}, {Amard}, {Barros}, {Boisse},
  {Bonomo}, {Bruno}, {Courcol}, {Deleuil}, {Demangeon}, {D{\'\i}az}, {Guillot},
  {Havel}, {Montagnier}, {Rajpurohit}, {Rey}, \& {Santos}}]{Santerne}
{Santerne}, A., {Moutou}, C., {Tsantaki}, M., {et~al.} 2016, \aap, 587, A64,
  \dodoi{10.1051/0004-6361/201527329}

\bibitem[{{Santos} {et~al.}(2004){Santos}, {Israelian}, \&
  {Mayor}}]{Santos2004}
{Santos}, N.~C., {Israelian}, G., \& {Mayor}, M. 2004, \aap, 415, 1153,
  \dodoi{10.1051/0004-6361:20034469}

\bibitem[{{Sharma} {et~al.}(2011){Sharma}, {Bland-Hawthorn}, {Johnston}, \&
  {Binney}}]{Sharma2011}
{Sharma}, S., {Bland-Hawthorn}, J., {Johnston}, K.~V., \& {Binney}, J. 2011,
  \apj, 730, 3, \dodoi{10.1088/0004-637X/730/1/3}

\bibitem[{{Sozzetti} {et~al.}(2009){Sozzetti}, {Torres}, {Latham}, {Stefanik},
  {Korzennik}, {Boss}, {Carney}, \& {Laird}}]{sozzetti2009}
{Sozzetti}, A., {Torres}, G., {Latham}, D.~W., {et~al.} 2009, \apj, 697, 544,
  \dodoi{10.1088/0004-637X/697/1/544}

\bibitem[{{Stassun} {et~al.}(2018{\natexlab{a}}){Stassun}, {Oelkers}, {Pepper},
  {Paegert}, {De Lee}, {Torres}, {Latham}, {Charpinet}, {Dressing}, {Huber},
  {Kane}, {L{\'e}pine}, {Mann}, {Muirhead}, {Rojas-Ayala}, {Silvotti},
  {Fleming}, {Levine}, \& {Plavchan}}]{Stassun2018}
{Stassun}, K.~G., {Oelkers}, R.~J., {Pepper}, J., {et~al.} 2018{\natexlab{a}},
  \aj, 156, 102, \dodoi{10.3847/1538-3881/aad050}

\bibitem[{{Stassun} {et~al.}(2018{\natexlab{b}}){Stassun}, {Oelkers}, {Pepper},
  {Paegert}, {De Lee}, {Torres}, {Latham}, {Charpinet}, {Dressing}, {Huber},
  {Kane}, {L{\'e}pine}, {Mann}, {Muirhead}, {Rojas-Ayala}, {Silvotti},
  {Fleming}, {Levine}, \& {Plavchan}}]{Statsun2018}
---. 2018{\natexlab{b}}, \aj, 156, 102, \dodoi{10.3847/1538-3881/aad050}

\bibitem[{{Thorngren} {et~al.}(2019){Thorngren}, {Marley}, \&
  {Fortney}}]{Thorngren2019}
{Thorngren}, D.~P., {Marley}, M.~S., \& {Fortney}, J.~J. 2019, Research Notes
  of the American Astronomical Society, 3, 128,
  \dodoi{10.3847/2515-5172/ab4353}

\bibitem[{{Tokovinin} {et~al.}(2013){Tokovinin}, {Fischer}, {Bonati},
  {Giguere}, {Moore}, {Schwab}, {Spronck}, \& {Szymkowiak}}]{Tokovinin2013}
{Tokovinin}, A., {Fischer}, D.~A., {Bonati}, M., {et~al.} 2013, \pasp, 125,
  1336, \dodoi{10.1086/674012}

\bibitem[{{Tsiaras} {et~al.}(2016){Tsiaras}, {Waldmann}, {Rocchetto}, {Varley},
  {Morello}, {Damiano}, \& {Tinetti}}]{pylight}
{Tsiaras}, A., {Waldmann}, I.~P., {Rocchetto}, M., {et~al.} 2016,
  {pylightcurve: Exoplanet lightcurve model}.
\newblock \doeprint{1612.018}

\bibitem[{{Udry} \& {Santos}(2007)}]{Udry2007}
{Udry}, S., \& {Santos}, N.~C. 2007, \araa, 45, 397,
  \dodoi{10.1146/annurev.astro.45.051806.110529}

\bibitem[{{Wang} {et~al.}(2015){Wang}, {Fischer}, {Horch}, \&
  {Huang}}]{Wang2015}
{Wang}, J., {Fischer}, D.~A., {Horch}, E.~P., \& {Huang}, X. 2015, \apj, 799,
  229, \dodoi{10.1088/0004-637X/799/2/229}

\bibitem[{{Wright} {et~al.}(2012){Wright}, {Marcy}, {Howard}, {Johnson},
  {Morton}, \& {Fischer}}]{Wright2012}
{Wright}, J.~T., {Marcy}, G.~W., {Howard}, A.~W., {et~al.} 2012, \apj, 753,
  160, \dodoi{10.1088/0004-637X/753/2/160}

\bibitem[{{Wyatt} {et~al.}(2007){Wyatt}, {Clarke}, \& {Greaves}}]{Wyatt2007}
{Wyatt}, M.~C., {Clarke}, C.~J., \& {Greaves}, J.~S. 2007, \mnras, 380, 1737,
  \dodoi{10.1111/j.1365-2966.2007.12244.x}

\bibitem[{{Yasui} {et~al.}(2010){Yasui}, {Kobayashi}, {Tokunaga}, {Saito}, \&
  {Tokoku}}]{Yasui2010}
{Yasui}, C., {Kobayashi}, N., {Tokunaga}, A.~T., {Saito}, M., \& {Tokoku}, C.
  2010, \apjl, 723, L113, \dodoi{10.1088/2041-8205/723/1/L113}

\bibitem[{{Zechmeister} \& {K{\"u}rster}(2009)}]{zechmeister2009}
{Zechmeister}, M., \& {K{\"u}rster}, M. 2009, \aap, 496, 577,
  \dodoi{10.1051/0004-6361:200811296}

\bibitem[{{Zhou} {et~al.}(2017){Zhou}, {Bakos}, {Hartman}, {Latham}, {Torres},
  {Bhatti}, {Penev}, {Buchhave}, {Kov{\'a}cs}, {Bieryla}, {Quinn}, {Isaacson},
  {Fulton}, {Falco}, {Csubry}, {Everett}, {Szklenar}, {Esquerdo}, {Berlind},
  {Calkins}, {B{\'e}ky}, {Knox}, {Hinz}, {Horch}, {Hirsch}, {Howell}, {Noyes},
  {Marcy}, {de Val-Borro}, {L{\'a}z{\'a}r}, {Papp}, \& {S{\'a}ri}}]{Zhou2017}
{Zhou}, G., {Bakos}, G.~{\'A}., {Hartman}, J.~D., {et~al.} 2017, \aj, 153, 211,
  \dodoi{10.3847/1538-3881/aa674a}

\bibitem[{{Zhou} {et~al.}(2019){Zhou}, {Huang}, {Bakos}, {Hartman}, {Latham},
  {Quinn}, {Collins}, {Winn}, {Wong}, {Kov{\'a}cs}, {Csubry}, {Bhatti},
  {Penev}, {Bieryla}, {Esquerdo}, {Berlind}, {Calkins}, {de Val-Borro},
  {Noyes}, {L{\'a}z{\'a}r}, {Papp}, {S{\'a}ri}, {Kov{\'a}cs}, {Buchhave},
  {Szklenar}, {B{\'e}ky}, {Johnson}, {Cochran}, {Kniazev}, {Stassun}, {Fulton},
  {Shporer}, {Espinoza}, {Bayliss}, {Everett}, {Howell}, {Hellier}, {Anderson},
  {Collier Cameron}, {West}, {Brown}, {Schanche}, {Barkaoui}, {Pozuelos},
  {Gillon}, {Jehin}, {Benkhaldoun}, {Daassou}, {Ricker}, {Vanderspek},
  {Seager}, {Jenkins}, {Lissauer}, {Armstrong}, {Collins}, {Gan}, {Hart},
  {Horne}, {Kielkopf}, {Nielsen}, {Nishiumi}, {Narita}, {Palle}, {Relles},
  {Sefako}, {Tan}, {Davies}, {Goeke}, {Guerrero}, {Haworth}, \&
  {Villanueva}}]{Zhou2020}
{Zhou}, G., {Huang}, C.~X., {Bakos}, G.~{\'A}., {et~al.} 2019, \aj, 158, 141,
  \dodoi{10.3847/1538-3881/ab36b5}

\bibitem[{{Zink} {et~al.}(2020){Zink}, {Hardegree-Ullman}, {Christiansen},
  {Dressing}, {Crossfield}, {Petigura}, {Schlieder}, \& {Ciardi}}]{edi}
{Zink}, J.~K., {Hardegree-Ullman}, K.~K., {Christiansen}, J.~L., {et~al.} 2020,
  \aj, 159, 154, \dodoi{10.3847/1538-3881/ab7448}

\end{thebibliography}
\nocite{*}

\end{CJK*}
\end{document}